\documentclass[journal,10pt,twoside,twocolumn]{IEEEtran}

\IEEEoverridecommandlockouts

\usepackage{amsmath,amsfonts} 
\usepackage{amssymb, amsthm}
\usepackage{graphicx}
\usepackage{subfigure}
\usepackage{booktabs}
\usepackage{cite}
\usepackage{xcolor}
\usepackage{microtype}
\usepackage{balance}
\usepackage{mathrsfs}
\usepackage{xspace}
\usepackage{bm}
\usepackage{upgreek}
\usepackage{fancyref}
\usepackage{textcomp}
\usepackage{multirow}
\usepackage{stmaryrd}

\newcommand{\safemath}[2]{\newcommand{#1}{\ensuremath{#2}\xspace}}

\safemath{\bma}{\mathbf{a}}
\safemath{\bmb}{\mathbf{b}}
\safemath{\bmc}{\mathbf{c}}
\safemath{\bmd}{\mathbf{d}}
\safemath{\bme}{\mathbf{e}}
\safemath{\bmf}{\mathbf{f}}
\safemath{\bmg}{\mathbf{g}}
\safemath{\bmh}{\mathbf{h}}
\safemath{\bmi}{\mathbf{i}}
\safemath{\bmj}{\mathbf{j}}
\safemath{\bmk}{\mathbf{k}}
\safemath{\bml}{\mathbf{l}}
\safemath{\bmm}{\mathbf{m}}
\safemath{\bmn}{\mathbf{n}}
\safemath{\bmo}{\mathbf{o}}
\safemath{\bmp}{\mathbf{p}}
\safemath{\bmq}{\mathbf{q}}
\safemath{\bmr}{\mathbf{r}}
\safemath{\bms}{\mathbf{s}}
\safemath{\bmt}{\mathbf{t}}
\safemath{\bmu}{\mathbf{u}}
\safemath{\bmv}{\mathbf{v}}
\safemath{\bmw}{\mathbf{w}}
\safemath{\bmx}{\mathbf{x}}
\safemath{\bmy}{\mathbf{y}}
\safemath{\bmz}{\mathbf{z}}
\safemath{\bmzero}{\mathbf{0}}
\safemath{\bmone}{\mathbf{1}}

\bmdefine{\biad}{a}
\bmdefine{\bibd}{b}
\bmdefine{\bicd}{c}
\bmdefine{\bidd}{d}
\bmdefine{\bied}{e}
\bmdefine{\bifd}{f}
\bmdefine{\bigd}{g}
\bmdefine{\bihd}{h}
\bmdefine{\biid}{i}
\bmdefine{\bijd}{j}
\bmdefine{\bikd}{k}
\bmdefine{\bild}{l}
\bmdefine{\bimd}{m}
\bmdefine{\bind}{n}
\bmdefine{\biod}{o}
\bmdefine{\bipd}{p}
\bmdefine{\biqd}{q}
\bmdefine{\bird}{r}
\bmdefine{\bisd}{s}
\bmdefine{\bitd}{t}
\bmdefine{\biud}{u}
\bmdefine{\bivd}{v}
\bmdefine{\biwd}{w}
\bmdefine{\bixd}{x}
\bmdefine{\biyd}{y}
\bmdefine{\bizd}{z}

\bmdefine{\bixid}{\xi}
\bmdefine{\bilambdad}{\lambda}
\bmdefine{\bimud}{\mu}
\bmdefine{\bithetad}{\theta}
\bmdefine{\biphid}{\phi}
\bmdefine{\bideltad}{\delta}

\safemath{\bmia}{\biad}
\safemath{\bmib}{\bibd}
\safemath{\bmic}{\bicd}
\safemath{\bmid}{\bidd}
\safemath{\bmie}{\bied}
\safemath{\bmif}{\bifd}
\safemath{\bmig}{\bigd}
\safemath{\bmih}{\bihd}
\safemath{\bmii}{\biid}
\safemath{\bmij}{\bijd}
\safemath{\bmik}{\bikd}
\safemath{\bmil}{\bild}
\safemath{\bmim}{\bimd}
\safemath{\bmin}{\bind}
\safemath{\bmio}{\biod}
\safemath{\bmip}{\bipd}
\safemath{\bmiq}{\biqd}
\safemath{\bmir}{\bird}
\safemath{\bmis}{\bisd}
\safemath{\bmit}{\bitd}
\safemath{\bmiu}{\biud}
\safemath{\bmiv}{\bivd}
\safemath{\bmiw}{\biwd}
\safemath{\bmix}{\bixd}
\safemath{\bmiy}{\biyd}
\safemath{\bmiz}{\bizd}

\safemath{\bmxi}{\bixid}
\safemath{\bmlambda}{\bilambdad}
\safemath{\bmmu}{\bimud}
\safemath{\bmtheta}{\bithetad}
\safemath{\bmphi}{\biphid}
\safemath{\bmdelta}{\bideltad}

\safemath{\bA}{\mathbf{A}}
\safemath{\bB}{\mathbf{B}}
\safemath{\bC}{\mathbf{C}}
\safemath{\bD}{\mathbf{D}}
\safemath{\bE}{\mathbf{E}}
\safemath{\bF}{\mathbf{F}}
\safemath{\bG}{\mathbf{G}}
\safemath{\bH}{\mathbf{H}}
\safemath{\bI}{\mathbf{I}}
\safemath{\bJ}{\mathbf{J}}
\safemath{\bK}{\mathbf{K}}
\safemath{\bL}{\mathbf{L}}
\safemath{\bM}{\mathbf{M}}
\safemath{\bN}{\mathbf{N}}
\safemath{\bO}{\mathbf{O}}
\safemath{\bP}{\mathbf{P}}
\safemath{\bQ}{\mathbf{Q}}
\safemath{\bR}{\mathbf{R}}
\safemath{\bS}{\mathbf{S}}
\safemath{\bT}{\mathbf{T}}
\safemath{\bU}{\mathbf{U}}
\safemath{\bV}{\mathbf{V}}
\safemath{\bW}{\mathbf{W}}
\safemath{\bX}{\mathbf{X}}
\safemath{\bY}{\mathbf{Y}}
\safemath{\bZ}{\mathbf{Z}}

\safemath{\bZero}{\mathbf{0}}
\safemath{\bOne}{\mathbf{1}}
\safemath{\bDelta}{\mathbf{\Delta}}
\safemath{\bLambda}{\mathbf{\UpLambda}}
\safemath{\bPhi}{\mathbf{\Upphi}}
\safemath{\bSigma}{\mathbf{\Upsigma}}
\safemath{\bOmega}{\mathbf{\Upomega}}
\safemath{\bTheta}{\mathbf{\Uptheta}}

\bmdefine{\biAd}{A}
\bmdefine{\biBd}{B}
\bmdefine{\biCd}{C}
\bmdefine{\biDd}{D}
\bmdefine{\biEd}{E}
\bmdefine{\biFd}{F}
\bmdefine{\biGd}{G}
\bmdefine{\biHd}{H}
\bmdefine{\biId}{I}
\bmdefine{\biJd}{J}
\bmdefine{\biKd}{K}
\bmdefine{\biLd}{L}
\bmdefine{\biMd}{M}
\bmdefine{\biOd}{N}
\bmdefine{\biPd}{O}
\bmdefine{\biQd}{P}
\bmdefine{\biRd}{R}
\bmdefine{\biSd}{S}
\bmdefine{\biTd}{T}
\bmdefine{\biUd}{U}
\bmdefine{\biVd}{V}
\bmdefine{\biWd}{W}
\bmdefine{\biXd}{X}
\bmdefine{\biYd}{Y}
\bmdefine{\biZd}{Z}

\bmdefine{\biDelta}{\Delta}
\bmdefine{\biLambda}{\Lambda}
\bmdefine{\biPhi}{\Phi}
\bmdefine{\biSigma}{\Sigma}
\bmdefine{\biOmega}{\Omega}
\bmdefine{\biTheta}{\Theta}

\safemath{\bimA}{\biAd}
\safemath{\bimB}{\biBd}
\safemath{\bimC}{\biCd}
\safemath{\bimD}{\biDd}
\safemath{\bimE}{\biEd}
\safemath{\bimF}{\biFd}
\safemath{\bimG}{\biGd}
\safemath{\bimH}{\biHd}
\safemath{\bimI}{\biId}
\safemath{\bimJ}{\biJd}
\safemath{\bimK}{\biKd}
\safemath{\bimL}{\biLd}
\safemath{\bimM}{\biMd}
\safemath{\bimN}{\biNd}
\safemath{\bimO}{\biOd}
\safemath{\bimP}{\biPd}
\safemath{\bimQ}{\biQd}
\safemath{\bimR}{\biRd}
\safemath{\bimS}{\biSd}
\safemath{\bimT}{\biTd}
\safemath{\bimU}{\biUd}
\safemath{\bimV}{\biVd}
\safemath{\bimW}{\biWd}
\safemath{\bimX}{\biXd}
\safemath{\bimY}{\biYd}
\safemath{\bimZ}{\biZd}

\safemath{\bimDelta}{\biDelta}
\safemath{\bimLambda}{\biLambda}
\safemath{\bimPhi}{\biPhi}
\safemath{\bimSigma}{\biSigma}
\safemath{\bimOmega}{\biOmega}
\safemath{\bimTheta}{\biTheta}

\safemath{\setA}{\mathcal{A}}
\safemath{\setB}{\mathcal{B}}
\safemath{\setC}{\mathcal{C}}
\safemath{\setD}{\mathcal{D}}
\safemath{\setE}{\mathcal{E}}
\safemath{\setF}{\mathcal{F}}
\safemath{\setG}{\mathcal{G}}
\safemath{\setH}{\mathcal{H}}
\safemath{\setI}{\mathcal{I}}
\safemath{\setJ}{\mathcal{J}}
\safemath{\setK}{\mathcal{K}}
\safemath{\setL}{\mathcal{L}}
\safemath{\setM}{\mathcal{M}}
\safemath{\setN}{\mathcal{N}}
\safemath{\setO}{\mathcal{O}}
\safemath{\setP}{\mathcal{P}}
\safemath{\setQ}{\mathcal{Q}}
\safemath{\setR}{\mathcal{R}}
\safemath{\setS}{\mathcal{S}}
\safemath{\setT}{\mathcal{T}}
\safemath{\setU}{\mathcal{U}}
\safemath{\setV}{\mathcal{V}}
\safemath{\setW}{\mathcal{W}}
\safemath{\setX}{\mathcal{X}}
\safemath{\setY}{\mathcal{Y}}
\safemath{\setZ}{\mathcal{Z}}
\safemath{\emptySet}{\varnothing}

\safemath{\colA}{\mathscr{A}}
\safemath{\colB}{\mathscr{B}}
\safemath{\colC}{\mathscr{C}}
\safemath{\colD}{\mathscr{D}}
\safemath{\colE}{\mathscr{E}}
\safemath{\colF}{\mathscr{F}}
\safemath{\colG}{\mathscr{G}}
\safemath{\colH}{\mathscr{H}}
\safemath{\colI}{\mathscr{I}}
\safemath{\colJ}{\mathscr{J}}
\safemath{\colK}{\mathscr{K}}
\safemath{\colL}{\mathscr{L}}
\safemath{\colM}{\mathscr{M}}
\safemath{\colN}{\mathscr{N}}
\safemath{\colO}{\mathscr{O}}
\safemath{\colP}{\mathscr{P}}
\safemath{\colQ}{\mathscr{Q}}
\safemath{\colR}{\mathscr{R}}
\safemath{\colS}{\mathscr{S}}
\safemath{\colT}{\mathscr{T}}
\safemath{\colU}{\mathscr{U}}
\safemath{\colV}{\mathscr{V}}
\safemath{\colW}{\mathscr{W}}
\safemath{\colX}{\mathscr{X}}
\safemath{\colY}{\mathscr{Y}}
\safemath{\colZ}{\mathscr{Z}}

\safemath{\opA}{\mathbb{A}}
\safemath{\opB}{\mathbb{B}}
\safemath{\opC}{\mathbb{C}}
\safemath{\opD}{\mathbb{D}}
\safemath{\opE}{\mathbb{E}}
\safemath{\opF}{\mathbb{F}}
\safemath{\opG}{\mathbb{G}}
\safemath{\opH}{\mathbb{H}}
\safemath{\opI}{\mathbb{I}}
\safemath{\opJ}{\mathbb{J}}
\safemath{\opK}{\mathbb{K}}
\safemath{\opL}{\mathbb{L}}
\safemath{\opM}{\mathbb{M}}
\safemath{\opN}{\mathbb{N}}
\safemath{\opO}{\mathbb{O}}
\safemath{\opP}{\mathbb{P}}
\safemath{\opQ}{\mathbb{Q}}
\safemath{\opR}{\mathbb{R}}
\safemath{\opS}{\mathbb{S}}
\safemath{\opT}{\mathbb{T}}
\safemath{\opU}{\mathbb{U}}
\safemath{\opV}{\mathbb{V}}
\safemath{\opW}{\mathbb{W}}
\safemath{\opX}{\mathbb{X}}
\safemath{\opY}{\mathbb{Y}}
\safemath{\opZ}{\mathbb{Z}}
\safemath{\opZero}{\mathbb{O}}
\safemath{\identityop}{\opI}

\safemath{\veca}{\bma}
\safemath{\vecb}{\bmb}
\safemath{\vecc}{\bmc}
\safemath{\vecd}{\bmd}
\safemath{\vece}{\bme}
\safemath{\vecf}{\bmf}
\safemath{\vecg}{\bmg}
\safemath{\vech}{\bmh}
\safemath{\veci}{\bmi}
\safemath{\vecj}{\bmj}
\safemath{\veck}{\bmk}
\safemath{\vecl}{\bml}
\safemath{\vecm}{\bmm}
\safemath{\vecn}{\bmn}
\safemath{\veco}{\bmo}
\safemath{\vecp}{\bmp}
\safemath{\vecq}{\bmq}
\safemath{\vecr}{\bmr}
\safemath{\vecs}{\bms}
\safemath{\vect}{\bmt}
\safemath{\vecu}{\bmu}
\safemath{\vecv}{\bmv}
\safemath{\vecw}{\bmw}
\safemath{\vecx}{\bmx}
\safemath{\vecy}{\bmy}
\safemath{\vecz}{\bmz}

\safemath{\veczero}{\bmzero}
\safemath{\vecone}{\bmone}
\safemath{\vecxi}{\bmxi}
\safemath{\veclambda}{\bmlambda}
\safemath{\vecmu}{\bmmu}
\safemath{\vectheta}{\bmtheta}
\safemath{\vecphi}{\bmphi}
\safemath{\vecdelta}{\bmdelta}

\safemath{\matA}{\bA}
\safemath{\matB}{\bB}
\safemath{\matC}{\bC}
\safemath{\matD}{\bD}
\safemath{\matE}{\bE}
\safemath{\matF}{\bF}
\safemath{\matG}{\bG}
\safemath{\matH}{\bH}
\safemath{\matI}{\bI}
\safemath{\matJ}{\bJ}
\safemath{\matK}{\bK}
\safemath{\matL}{\bL}
\safemath{\matM}{\bM}
\safemath{\matN}{\bN}
\safemath{\matO}{\bO}
\safemath{\matP}{\bP}
\safemath{\matQ}{\bQ}
\safemath{\matR}{\bR}
\safemath{\matS}{\bS}
\safemath{\matT}{\bT}
\safemath{\matU}{\bU}
\safemath{\matV}{\bV}
\safemath{\matW}{\bW}
\safemath{\matX}{\bX}
\safemath{\matY}{\bY}
\safemath{\matZ}{\bZ}
\safemath{\matzero}{\bmzero}

\safemath{\matDelta}{\bDelta}
\safemath{\matLambda}{\bLambda}
\safemath{\matPhi}{\bPhi}
\safemath{\matSigma}{\bSigma}
\safemath{\matOmega}{\bOmega}
\safemath{\matTheta}{\bTheta}

\safemath{\matidentity}{\matI}
\safemath{\matone}{\matO}

\safemath{\rnda}{A}
\safemath{\rndb}{B}
\safemath{\rndc}{C}
\safemath{\rndd}{D}
\safemath{\rnde}{E}
\safemath{\rndf}{F}
\safemath{\rndg}{G}
\safemath{\rndh}{H}
\safemath{\rndi}{I}
\safemath{\rndj}{J}
\safemath{\rndk}{K}
\safemath{\rndl}{L}
\safemath{\rndm}{M}
\safemath{\rndn}{N}
\safemath{\rndo}{O}
\safemath{\rndp}{P}
\safemath{\rndq}{Q}
\safemath{\rndr}{R}
\safemath{\rnds}{S}
\safemath{\rndt}{T}
\safemath{\rndu}{U}
\safemath{\rndv}{V}
\safemath{\rndw}{W}
\safemath{\rndx}{X}
\safemath{\rndy}{Y}
\safemath{\rndz}{Z}

\safemath{\rveca}{\bimA}
\safemath{\rvecb}{\bimB}
\safemath{\rvecc}{\bimC}
\safemath{\rvecd}{\bimD}
\safemath{\rvece}{\bimE}
\safemath{\rvecf}{\bimF}
\safemath{\rvecg}{\bimG}
\safemath{\rvech}{\bimH}
\safemath{\rveci}{\bimI}
\safemath{\rvecj}{\bimJ}
\safemath{\rveck}{\bimK}
\safemath{\rvecl}{\bimL}
\safemath{\rvecm}{\bimM}
\safemath{\rvecn}{\bimN}
\safemath{\rveco}{\bomO}
\safemath{\rvecp}{\bimP}
\safemath{\rvecq}{\bimQ}
\safemath{\rvecr}{\bimR}
\safemath{\rvecs}{\bimS}
\safemath{\rvect}{\bimT}
\safemath{\rvecu}{\bimU}
\safemath{\rvecv}{\bimV}
\safemath{\rvecw}{\bimW}
\safemath{\rvecx}{\bimX}
\safemath{\rvecy}{\bimY}
\safemath{\rvecz}{\bimZ}

\safemath{\rvecxi}{\bmxi}
\safemath{\rveclambda}{\bmlambda}
\safemath{\rvecmu}{\bmmu}
\safemath{\rvectheta}{\bmtheta}
\safemath{\rvecphi}{\bmphi}

\safemath{\rmatA}{\bimA}
\safemath{\rmatB}{\bimB}
\safemath{\rmatC}{\bimC}
\safemath{\rmatD}{\bimD}
\safemath{\rmatE}{\bimE}
\safemath{\rmatF}{\bimF}
\safemath{\rmatG}{\bimG}
\safemath{\rmatH}{\bimH}
\safemath{\rmatI}{\bimI}
\safemath{\rmatJ}{\bimJ}
\safemath{\rmatK}{\bimK}
\safemath{\rmatL}{\bimL}
\safemath{\rmatM}{\bimM}
\safemath{\rmatN}{\bimN}
\safemath{\rmatO}{\bimO}
\safemath{\rmatP}{\bimP}
\safemath{\rmatQ}{\bimQ}
\safemath{\rmatR}{\bimR}
\safemath{\rmatS}{\bimS}
\safemath{\rmatT}{\bimT}
\safemath{\rmatU}{\bimU}
\safemath{\rmatV}{\bimV}
\safemath{\rmatW}{\bimW}
\safemath{\rmatX}{\bimX}
\safemath{\rmatY}{\bimY}
\safemath{\rmatZ}{\bimZ}

\safemath{\rmatDelta}{\bimDelta}
\safemath{\rmatLambda}{\bimLambda}
\safemath{\rmatPhi}{\bimPhi}
\safemath{\rmatSigma}{\bimSigma}
\safemath{\rmatOmega}{\bimOmega}
\safemath{\rmatTheta}{\bimTheta}

\newenvironment{textbmatrix}{	\setlength{\arraycolsep}{2.5pt}%
								\big[\begin{matrix}}{\end{matrix}\big]%
								\raisebox{0.08ex}{\vphantom{M}}}

\def\be{\begin{equation}}
\def\ee{\end{equation}}
\def\een{\nonumber \end{equation}}
\def\mat{\begin{bmatrix}}
\def\emat{\end{bmatrix}}
\def\btm{\begin{textbmatrix}}
\def\etm{\end{textbmatrix}}

\def\ba#1\ea{\begin{align}#1\end{align}}
\def\bas#1\eas{\begin{align*}#1\end{align*}}
\def\bs#1\es{\begin{split}#1\end{split}} 
\def\bg#1\eg{\begin{gather}#1\end{gather}}
\def\bml#1\eml{\begin{multline}#1\end{multline}}
\def\bi#1\ei{\begin{itemize}#1\end{itemize}}

\newcommand{\lefto}{\mathopen{}\left}

\DeclareMathOperator{\Exop}{\opE}			

\newcommand{\Ex}[2]{\ensuremath{\Exop_{#1}\lefto[#2\right]}} 	

\newcommand{\vecnorm}[1]{\lefto\lVert#1\right\rVert}		

\safemath{\dirac}{\delta}					
\safemath{\krond}{\dirac}					

\safemath{\upto}{\uparrow}
\safemath{\downto}{\downarrow}
\safemath{\iu}{j}							
\safemath{\ev}{\lambda}						
\safemath{\hilseqspace}{l^{2}}				
\newcommand{\banachfunspace}[1]{\setL^{#1}}	
\safemath{\hilfunspace}{\banachfunspace{2}}	

\safemath{\SNR}{\textsf{SNR}} 				
\safemath{\PAR}{\textsf{PAR}} 				
\safemath{\No}{N_0}							
\safemath{\Es}{E_s}							
\safemath{\Eb}{E_b}							
\safemath{\EbNo}{\frac{\Eb}{\No}}
\safemath{\EsNo}{\frac{\Es}{\No}}

\DeclareMathOperator{\CHop}{\ensuremath{\opH}} 
\safemath{\tvir}{\rndh_{\CHop}}				
\safemath{\tvtf}{\rndl_{\CHop}}				
\safemath{\spf}{\rnds_{\CHop}}				
\safemath{\bff}{H_{\CHop}}					

\safemath{\ircf}{r_{h}}						
\safemath{\tftvcf}{r_{s}}					
\safemath{\tfcf}{r_{l}}						
\safemath{\bfcf}{r_{H}}						

\safemath{\tcorr}{c_h}						
\safemath{\scf}{c_{s}}						
\safemath{\tfcorr}{c_{l}}					
\safemath{\fcorr}{c_{H}}						

\safemath{\mi}{I}							
\safemath{\capacity}{C}						

\safemath{\normal}{\mathcal{N}}			
\safemath{\jpg}{\mathcal{CN}}			
\safemath{\mchain}{\leftrightarrow}		

\safemath{\dB}{\,\mathrm{dB}}
\safemath{\dBm}{\,\mathrm{dBm}}
\safemath{\Hz}{\,\mathrm{Hz}}
\safemath{\kHz}{\,\mathrm{kHz}}
\safemath{\MHz}{\,\mathrm{MHz}}
\safemath{\GHz}{\,\mathrm{GHz}}
\safemath{\s}{\,\mathrm{s}}
\safemath{\ms}{\,\mathrm{ms}}
\safemath{\mus}{\,\mathrm{\text{\textmu}s}}
\safemath{\ns}{\,\mathrm{ns}}
\safemath{\ps}{\,\mathrm{ps}}
\safemath{\meter}{\,\mathrm{m}}
\safemath{\mm}{\,\mathrm{mm}}
\safemath{\cm}{\,\mathrm{cm}}
\safemath{\m}{\,\mathrm{m}}
\safemath{\W}{\,\mathrm{W}}
\safemath{\mW}{\, \mathrm{mW}}
\safemath{\J}{\,\mathrm{J}}
\safemath{\K}{\,\mathrm{K}}
\safemath{\bit}{\,\mathrm{bit}}
\safemath{\nat}{\,\mathrm{nat}}

\safemath{\define}{\triangleq}			

\safemath{\equivalent}{\sim}
\safemath{\distas}{\sim}					
\safemath{\sdiff}{\Delta}				

\safemath{\reals}{\mathbb{R}}
\safemath{\positivereals}{\reals_{+}}
\safemath{\integers}{\mathbb{Z}}
\safemath{\posint}{\integers_{+}}
\safemath{\naturals}{\mathbb{N}}
\safemath{\posnaturals}{\naturals_{+}}
\safemath{\complexset}{\mathbb{C}}
\safemath{\rationals}{\mathbb{Q}}

\newcommand*{\fancyrefapplabelprefix}{app}		
\newcommand*{\fancyrefthmlabelprefix}{thm}		
\newcommand*{\fancyreflemlabelprefix}{lem}		
\newcommand*{\fancyrefcorlabelprefix}{cor}		
\newcommand*{\fancyrefdeflabelprefix}{def}		
\newcommand*{\fancyrefproplabelprefix}{prop}	
\newcommand*{\fancyrefobslabelprefix}{obs}		
\newcommand*{\fancyrefalglabelprefix}{alg}		
\newcommand*{\fancyrefasmlabelprefix}{asm}	    
\newcommand*{\fancyreftbllabelprefix}{tbl}	    

\frefformat{vario}{\fancyrefseclabelprefix}{Section~#1}
\frefformat{vario}{\fancyrefthmlabelprefix}{Theorem~#1}
\frefformat{vario}{\fancyreflemlabelprefix}{Lemma~#1}
\frefformat{vario}{\fancyrefcorlabelprefix}{Corollary~#1}
\frefformat{vario}{\fancyrefdeflabelprefix}{Definition~#1}
\frefformat{vario}{\fancyrefobslabelprefix}{Observation~#1}
\frefformat{vario}{\fancyrefasmlabelprefix}{Assumption~#1}
\frefformat{vario}{\fancyreffiglabelprefix}{Figure~#1}
\frefformat{vario}{\fancyrefapplabelprefix}{Appendix~#1} 
\frefformat{vario}{\fancyrefproplabelprefix}{Proposition~#1}
\frefformat{vario}{\fancyrefalglabelprefix}{Algorithm~#1}
\frefformat{vario}{\fancyrefeqlabelprefix}{(#1)}
\frefformat{vario}{\fancyreftbllabelprefix}{Table~#1}

\safemath{\dictab}{[\,\dicta\,\,\dictb\,]}

\safemath{\ysig}{\bmy}
\safemath{\ysighat}{\hat{\ysig}}
\safemath{\ysigdim}{M}
\safemath{\xsig}{\bmx}
\safemath{\xsigdim}{N}
\safemath{\nx}{n_x}
\safemath{\zsig}{\bmz}
\safemath{\zsigdim}{\ysigdim}
\safemath{\rsig}{\bmr}
\safemath{\Adict}{\bA}
\safemath{\Adicttilde}{\widetilde{\Adict}}
\safemath{\Adictdim}{\outputdim\times\xsigdim}
\safemath{\avec}{\bma}
\safemath{\avectilde}{\tilde{\avec}}
\safemath{\Bdict}{\bB}
\safemath{\Bdicttilde}{\widetilde{\Bdict}}
\safemath{\Cdict}{\bC}
\safemath{\cvec}{\bmc}
\safemath{\Ddict}{\bD}
\safemath{\Ddictdim}{\ysigdim\times\xsigdim}
\safemath{\dvec}{\bmd}
\safemath{\Ddicttilde}{\widetilde{\bD}}
\safemath{\Bonb}{\bB}
\safemath{\bvec}{\bmb}
\safemath{\Bonbdim}{\ysigdim\times\ysigdim}
\safemath{\noise}{\bmn}
\safemath{\noisedim}{\ysigim}
\safemath{\err}{\bme}
\safemath{\errdim}{\ysigdim}
\safemath{\errset}{\setE}
\safemath{\nerr}{n_e}
\safemath{\delop}{\bP_\errset}
\safemath{\delopc}{\bP_{{\errset}^c}}

\safemath{\cplxi}{\imath}
\safemath{\cplxj}{\jmath}

\safemath{\dict}{\matD}
\safemath{\inputdim}{N}		
\safemath{\outputdim}{M}		
\safemath{\sparsity}{S}	
\safemath{\inputdimA}{{N_a}}	
\safemath{\inputdimB}{{N_b}}	
\safemath{\elemA}{{n_a}}	
\safemath{\elemB}{{n_b}}	
\safemath{\resA}{\matR_a}	
\safemath{\resB}{\matR_b}	
\safemath{\subD}{\matS} 
\safemath{\subA}{\matS_a} 
\safemath{\subB}{\matS_b} 
\safemath{\dicta}{\matA} 	
\safemath{\dictb}{\matB} 	
\safemath{\hollowS}{H}
\safemath{\hollowA}{H_a}
\safemath{\hollowB}{H_b}
\safemath{\cross}{Z}
\safemath{\coh}{\mu_d}			
\safemath{\coha}{\mu_a}			
\safemath{\cohb}{\mu_b}			
\safemath{\mubs}{\nu}	
\safemath{\cohm}{\mu_m} 
\safemath{\dictset}{\setD}	
\safemath{\dictsetp}{\dictset(\coh,\coha,\cohb)}	
\safemath{\dictsetgen}{\dictset_\text{gen}}
\safemath{\dictsetgenp}{\dictsetgen(\coh)}
\safemath{\dictsetonb}{\dictset_\text{onb}}
\safemath{\dictsetonbp}{\dictsetonb(\coh)}

\safemath{\leftside}{U}
\safemath{\rightsideA}{R_a}
\safemath{\rightsideB}{R_b}

\safemath{\indexS}{\setI_S} 

\safemath{\na}{n_a}			
\safemath{\nb}{n_b}			
\safemath{\coeffa}{p_i}	
\safemath{\coeffb}{q_j}	
\safemath{\seta}{\setP}		
\safemath{\setb}{\setQ}     
\safemath{\setw}{\setW}	
\safemath{\setz}{\setZ}	
\safemath{\cola}{\veca}		
\safemath{\colb}{\vecb}		
\safemath{\cold}{\vecd}		
\safemath{\inputvec}{\vecx} 	
\safemath{\error}{\vece}	
\safemath{\noiseout}{\vecz} 	
\safemath{\inputvecel}{x}
\safemath{\inputveca}{\vecx_a}
\safemath{\inputvecb}{\vecx_b}
\safemath{\outputvec}{\vecy}	
\safemath{\lambdamin}{\lambda_{\mathrm{min}}}

\safemath{\elltwo}{\ell_2}
\safemath{\ellone}{\ell_1}
\safemath{\ellzero}{\ell_0}
\safemath{\ellinf}{\ell_\infty}
\safemath{\ellinftilde}{\ell_{\widetilde\infty}}
\safemath{\licard}{Z(\coh,\coha,\cohb)}
\safemath{\xsol}{\hat{x}}
\safemath{\xbord}{x_b}		
\safemath{\xstat}{x_s}		
\safemath{\xstatLone}{\tilde{x}_s}
\safemath{\order}{\mathcal{O}} 
\safemath{\scales}{\Theta} 
\safemath{\ones}{\mathbf{1}} 
\safemath{\zeroes}{\mathbf{0}} 
\safemath{\thlone}{\kappa(\coh,\cohb)} 
\safemath{\constoneA}{\delta} 
\safemath{\constoneB}{\epsilon} 
\safemath{\nlarge}{L}				   
\safemath{\sumlarge}{S_\nlarge}
\safemath{\maxlarger}{P_\nlarge}	   
\safemath{\Pzero}{\textrm{P0}}	
\safemath{\Pone}{\textrm{P1}}
\safemath{\vecfir}{\vecw}			 
\safemath{\vecsec}{\vecz}
\safemath{\elvecfir}{w}              
\safemath{\elvecsec}{z}				 
\safemath{\nlargefir}{n}
\safemath{\normout}{\gamma}
\safemath{\auxfun}{h}
\safemath{\supp}{\textrm{supp}}

\safemath{\indexa}{\ell}
\safemath{\indexb}{r}
\safemath{\indexc}{i}
\safemath{\indexd}{j}

\safemath{\project}{P}

\renewcommand{\bml}{\ensuremath{\boldsymbol \ell}}

\begin{document}
\title{High-Bandwidth Spatial Equalization for mmWave Massive MU-MIMO with Processing-In-Memory}
\author{
\IEEEauthorblockN{Oscar Casta\~neda, Sven Jacobsson, Giuseppe Durisi, Tom Goldstein, and Christoph Studer}
\thanks{The work of O. Casta\~neda was supported by ComSenTer, a Semiconductor Research Corporation (SRC) program, by SRC nCORE task 2758.004, and by a Qualcomm Innovation Fellowship. The work of S. Jacobsson and G. Durisi was supported by the Swedish Foundation for Strategic Research under grants SM13-0028 and ID14-0022, and by the Swedish Governmental Agency for Innovation Systems (VINNOVA) within the VINN Excellence center Chase. The work of T. Goldstein was supported in part by the US National Science Foundation (NSF) under grant CCF-1535902 and by the US Office of Naval Research under grant N00014-15-1-2676. The work of C. Studer was supported  by Xilinx Inc.\ and by the US NSF under grants CCF-1652065, CNS-1717559, and ECCS-1824379.}
\thanks{O. Casta\~neda and C. Studer are with the School~of Electrical and Computer Engineering, Cornell Tech, New York, NY; e-mail: oc66@cornell.edu, studer@cornell.edu; web: http://vip.ece.cornell.edu} 
\thanks{S. Jacobsson is with Ericsson Research, Gothenburg, Sweden, and also with Chalmers University of Technology, Gothenburg, Sweden; e-mail: sven.jacobsson@ericsson.com}
\thanks{G. Durisi is with Chalmers University of Technology, Gothenburg, Sweden; e-mail: durisi@chalmers.se}
\thanks{T. Goldstein is with the Department of Computer Science, University of Maryland, College Park, MD; e-mail: tomg@cs.umd.edu}
}
\maketitle

\begin{abstract} 
All-digital basestation (BS) architectures enable superior spectral efficiency compared to hybrid solutions in massive multi-user MIMO systems. However, supporting large bandwidths with all-digital architectures at mmWave frequencies is challenging as traditional baseband processing would result in excessively high power consumption and large silicon area. The recently-proposed concept of finite-alphabet equalization is able to address both of these issues by using equalization matrices that contain low-resolution entries to lower the power and complexity of high-throughput matrix-vector products in hardware. In this paper, we explore two different finite-alphabet equalization hardware implementations that tightly integrate the memory and processing elements: (i) a parallel array of multiply-accumulate (MAC) units and (ii) a bit-serial processing-in-memory (PIM) architecture. Our all-digital VLSI implementation results in 28nm CMOS show that the bit-serial PIM architecture reduces the area and power consumption up to a factor of $\bf2\boldsymbol\times$ and $\bf3\boldsymbol\times$, respectively, when compared to a parallel MAC array that operates at the same throughput. 
\end{abstract}

\section{Introduction}
Future wireless systems are expected to rely on millimeter wave (mmWave) communication~\cite{swindlehurst14a} that provides extreme bandwidths, and massive multi-user multiple-input multiple-output (MU-MIMO) \cite{larsson14a} that compensates for the path loss at mmWave frequencies and enables communication with multiple user equipments (UEs) in the same time-frequency resource. 
However, the combination of high-bandwidth mmWave communication with hundreds of basestation (BS) antenna elements inevitably results in excessively high baseband complexity and power consumption.
In order to develop power-efficient BS designs for such systems, significant attention has been given to hybrid analog-digital architectures~\cite{roh2014millimeter,sadhu20177,du2018hybrid,magueta2019hybrid,alkhateeb14b}.
However, such hybrid solutions are limited in the number of transmission paths they can  resolve  simultaneously~\cite{alkhateeb14b,bjornson2019massive,dutta2019case}, thus degrading spectral efficiency.
All-digital BS architectures are a promising alternative to mitigate this drawback~\cite{mo15b,roth2017achievable,jacobsson17b}.
While it is widely believed that all-digital BS architectures consume more power than hybrid solutions, recent results in~\cite{dutta2019case,roth2017achievable} indicate that the power consumption of radio-frequency (RF) and data converters in all-digital architectures is comparable to that of hybrid systems when reducing the data-converter resolution.
Despite these recent findings,  the power consumption and system costs of baseband processing for all-digital mmWave massive MU-MIMO architectures are largely unexplored.

\subsection{All-Digital Spatial Equalization}
In the uplink, $U$ UEs transmit information to a BS equipped with $B$ antennas.
To recover the transmitted signals, the BS must perform spatial equalization for each received sample. For linear equalization methods, one has to compute complex-valued matrix-vector products at the rate of the baseband analog-to-digital converters (ADCs). 
In mmWave massive MU-MIMO systems, even such straightforward matrix-vector products will result in power-hungry digital circuitry as we are dealing with extremely high sampling rates and large equalization matrices. 
For example, a conventional digital circuit that computes matrix-vector products for $B=256$~antennas and $U=16$~UEs at a rate of $2$\,G\,vectors/s  consumes $28$\,W and occupies $129\,\text{mm}^2$ in $28$\,nm CMOS~\cite{castaneda19fame}.
The power and area will further increase when considering  systems with more BS antennas, more UEs, and higher sampling rates.
Consequently, all-digital BS architectures require efficient spatial equalization circuitry that minimizes power and area without degrading spectral~efficiency.

The hardware complexity (in terms of power and area) of matrix-vector products can be reduced by decreasing the number of bits used to represent its operands.
Previous work has focused extensively on reducing the received vector's precision, which corresponds to the use of low-resolution ADCs  (e.g., $1$ to $8$ bits)  at the BS of massive MU-MIMO systems~\cite{alkhateeb14b,dutta2019case,mo15b,roth2017achievable,studer16a,yan2019performance}.
The spatial equalization matrices, however, are typically represented using high-precision numbers (e.g., $10$ to $12$ bits)~\cite{studer2011asic,wu2014large}.
Recently, reference~\cite{castaneda19fame} proposed \emph{finite-alphabet equalization}, which represents equalization matrices with low-resolution numbers while minimizing the post-equalization mean-square error (MSE).
In~\cite{castaneda19fame}, finite-alphabet equalization was shown to reduce the equalization power and area by $3.9\times$ and $5.8\times$, respectively, when using conventional digital very-large scale integration (VLSI) designs.
\subsection{Processing-In-Memory (PIM)}
While the performance of digital VLSI designs continuously increased over many decades, memory access times have not improved at the same pace. This disparity led to a so-called ``memory-wall'' \cite{wulf1995hitting} in which communication with memories causes a major bottleneck in terms of throughput and energy efficiency.
\emph{Processing-in-memory} (PIM) is an emerging compute paradigm that aims at avoiding the memory wall by co-locating logic close to memories, with the goal of minimizing time and energy required for data movement~\cite{nair2015evolution}.
With the looming end of Moore's Law, PIM has caught increasing attention. Existing PIM approaches focus on incorporating logic into memory processes \cite{li2017drisa}, exploiting mixed-signal techniques~\cite{jia2018microprocessor}, using emerging devices \cite{guo2013ac}, or relying on standard CMOS logic processes with all-digital processing~\cite{castaneda2019ppac}.

\vspace{-0.15cm}
\subsection{Contributions}
Traditional application-specific integrated circuit (ASIC) designs are closely related to PIM, as  design-specific memory structures are placed near computation elements in order to maximize throughput and energy efficiency.
However, with the recent appearance of numerous PIM architectures, it is natural to ask whether PIM is useful for next-generation wireless systems, a domain which has largely benefitted from ASIC designs in the past.
To shed light on this question, we evaluate two distinct VLSI designs that implement finite-alphabet equalization.
The first design corresponds to an array of all-digital multiply-accumulate (MAC) units, which represents traditional ASICs.
The second design corresponds to a PIM approach, which equips the bit-cells of a memory array with XNOR functionality to enable efficient, massively-parallel low-resolution matrix-vector products.
To enable a fair comparison between PIM and traditional ASIC designs, we use a specialized version of the recently-proposed, all-digital PPAC, which stands for Parallel Processor in Associative Content addressable memory (CAM)~\cite{castaneda2019ppac}.
For the same finite-alphabet equalization throughput, we compare both solutions in terms of area and power consumption for a $28$\,nm CMOS technology.

\vspace{-0.15cm}
\subsection{Notation}
Uppercase bold letters denote matrices; lowercase, column vectors.
For a matrix $\bA$, the transpose, Hermitian transpose, real and imaginary parts are $\bA^T$, $\bA^H$, $\Re\{\bA\}$, and $\Im\{\bA\}$, respectively. 
For a vector~$\bma$, the $k$th entry is $a_k$, the  $\ell_2$-norm is $\vecnorm{\veca}_2$, and the entry-wise complex conjugate is $\bma^*$. 
$\Ex{\bmx}{\cdot}$ is the expectation operator with respect to the random vector $\bmx$.
%
\section{Spatial Equalization}
\subsection{System Model and Spatial Equalization Basics}
We consider the uplink of a narrowband\footnote{Our analysis is also suitable for wideband channels in combination with orthogonal frequency-division multiplexing (OFDM), where the input-output relation of each subcarrier can also be modeled as in \fref{eq:channelmodel}.} mmWave massive MU-MIMO system where a $B$-antenna BS receives signals from $U$ single-antenna UEs.
The uplink narrowband input-output relation is modeled as
\begin{align} \label{eq:channelmodel}
\bmy=\bH\bms+\bmn,
\end{align}
where $\bmy\in\complexset^B$ is the received vector at the BS,  $\bH\in\complexset^{B\times U}$ is the known uplink MIMO channel matrix, $\bms\in\setS^U$ is the transmit data vector, with $\setS$ being the constellation set (e.g., 16-QAM),  and  $\bmn\in\complexset^B$ is i.i.d.\ circularly-symmetric complex Gaussian noise with variance $\No$ per entry.
We assume that the covariance matrix of $\bms$ is $\bC_\bms=\Ex{\bms}{\bms\bms^H}=\Es\bI_U$.
Furthermore, we assume perfect channel state information at the BS.

Spatial equalization produces an estimate $\hat\bms \in \opC^U$ of the transmit data vector $\bms$ given $\bmy$ and $\bH$.
With linear equalization, the estimate $\hat\bms$ can be computed as 
\begin{align}
\hat\bms=\bW^H\bmy,
\end{align}
with the spatial equalization matrix $\bW^H\in\complexset^{U\times B}$.
Typically, $\bW^H$ is chosen to minimize the MSE defined as
\begin{align}\label{eq:mse}
\Ex{\bms,\bmn}{\|\bW^H\bmy-\bms\|^2_2}\!,
\end{align}
which results in the linear minimum MSE (L-MMSE) equalizer, given by \cite{paulraj03}
\begin{align}\label{eq:lmmse}
\bW^H = (\bH^H\bH+\rho\bI_U)^{-1}\bH^H
\end{align}
with $\rho=\No/\Es$. The complex-valued entries of the \mbox{L-MMSE} equalizer are routinely represented with high-resolution numbers (e.g., $10$ to $12$ bits \cite{studer2011asic,wu2014large}). For high-bandwidth mmWave massive MU-MIMO systems, such resolution leads to excessively large and power-hungry VLSI circuits.

\begin{figure}[tp]
\centering
\includegraphics[width=0.75\columnwidth]{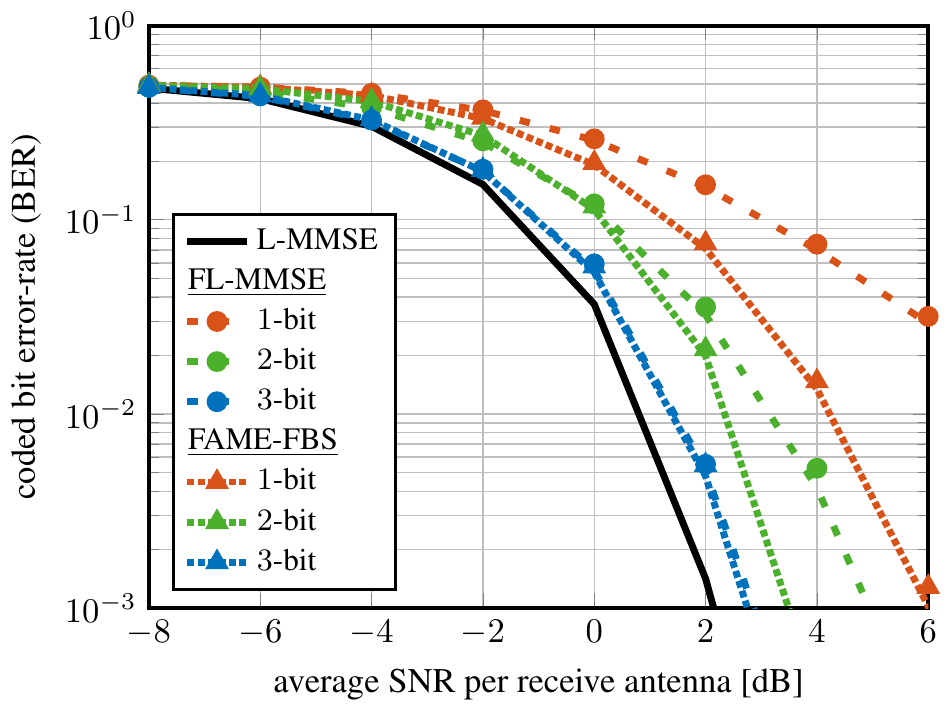}\\
\vspace{-0.2cm}
\caption{Bit error-rate (BER) for a $B=256$ BS antenna, $U=16$ UE, $16$-QAM, rate-$3/4$ coded OFDM mmWave MU-MIMO system operating under non-line-of-sight conditions. The curves represent the performance obtained when computing $\bW^H\bmy$ using double-precision floating-point arithmetic; the markers represent the performance obtained when computing $\bW^H\bmy$ using the fixed-point hardware design proposed in \fref{sec:ppac}. $1$-bit FAME-FBS significantly outperforms $1$-bit FL-MMSE. $3$-bit FAME-FBS and FL-MMSE exhibit similar performance and approach the performance of the infinite-precision L-MMSE. Figure adopted from \cite[Fig.~2(b)]{castaneda20a}.}
\label{fig:ber}
\end{figure}

\subsection{Finite-Alphabet Equalization}
To arrive at more efficient equalization circuitry, finite-alphabet equalization, put forward in~\cite{castaneda19fame}, proposes to use \emph{finite-alphabet equalization matrices}  with the following structure:
\begin{align} \label{eq:FAMEmatrix}
\bV^H = \mathrm{diag}(\boldsymbol\beta^*) \bX^H.
\end{align}
Here, $\bX^H\in\setX^{U\times B}$ is a low-resolution matrix whose entries come from a low-cardinality finite alphabet $\setX$ (e.g., the ``1-bit'' alphabet is $\{\pm1\pm j\}$), and $\boldsymbol\beta\in\complexset^U$ contains per-UE high-resolution scaling factors.
The work in~\cite{castaneda19fame} proposes two ways to compute a finite-alphabet matrix with the form in~\fref{eq:FAMEmatrix}: (i) quantizing the L-MMSE matrix in \fref{eq:lmmse} to a finite-alphabet, called FL-MMSE, and (ii) approximately solving the finite-alphabet MMSE equalization (FAME) problem using forward-backward splitting, called FAME-FBS.
\fref{fig:ber}, which is adopted from \cite[Fig.~2(b)]{castaneda20a}, clearly demonstrates that FAME-FBS enables superior error-rate performance, even when considering transmission over realistic, QuaDRiGa-generated~\cite{jaeckel2014quadriga} mmWave non-line-of-sight channels.
We note that FAME-FBS outperforms FL-MMSE as it computes finite-alphabet equalization matrices that minimize the MSE in \fref{eq:mse}.

Regardless of how the finite-alphabet matrices are computed, the structure in \fref{eq:FAMEmatrix} enables efficient hardware implementations as per-UE equalization becomes 
\begin{align}
\hat{s}_u=\bmv_u^H\bmy=\beta^*_u(\bmx_u^H\bmy),
\end{align}
where $\bmv_u^H$ and $\bmx_u^H$ are the $u$th rows of $\bV^H$ and $\bX^H$, respectively.
Since $\bmx_u^H$ has low-precision entries, the inner product $\bmx_u^H\bmy$ (requiring $B$ complex-valued scalar multiplications) can be computed efficiently using low-precision circuitry (e.g., adders and subtractors for the 1-bit case).
The low-resolution inner product is then scaled by $\beta^*_u$, a scalar operation that is carried out at higher resolution, but only once per UE.
\section{VLSI Architectures}
\label{sec:arch}
We will now detail three different VLSI architectures that implement finite-alphabet equalization.
All of these architectures tightly integrate the datapath and memory to achieve high throughput and energy efficiency.
The first and second architectures achieve such integration with a traditional ASIC design approach---the third one relies on all-digital PIM.

We consider VLSI architectures that perform matrix-vector multiplications $\bV^H\bmy$ with a finite-alphabet equalization matrix $\bV^H$.
Furthermore, we assume that the low-resolution part~$\bX^H$ of such matrix $\bV^H$ is represented with a symmetric set of mid-rise quantized numbers, e.g., 
the $2$-bit alphabet $\setX$ has entries whose real and imaginary parts are in the set $\{\pm1,\pm3\}$. We represent the entries of $\bmy$ using two's complement numbers.

\subsection{Linear Array of MAC Units}
As a baseline, we consider the architecture in~\cite{castaneda19fame}, which corresponds to a linear array of $U$ complex-valued MAC units (one per UE).
Each MAC unit reads data from a memory storing the corresponding $B$-dimensional row $\bmx_u^H$ of~$\bX^H$ and all MAC units receive one entry of $\bmy$ per clock cycle.
Thus, the inner product $\bmx_u^H\bmy$ is computed in $B$ clock cycles.
The result is then scaled by $\beta^*_u$ using a high-resolution multiplier.
\subsection{Optimized MAC Array}
\label{sec:repmac}
\begin{figure}[tp]
\centering
\includegraphics[width=0.76\columnwidth]{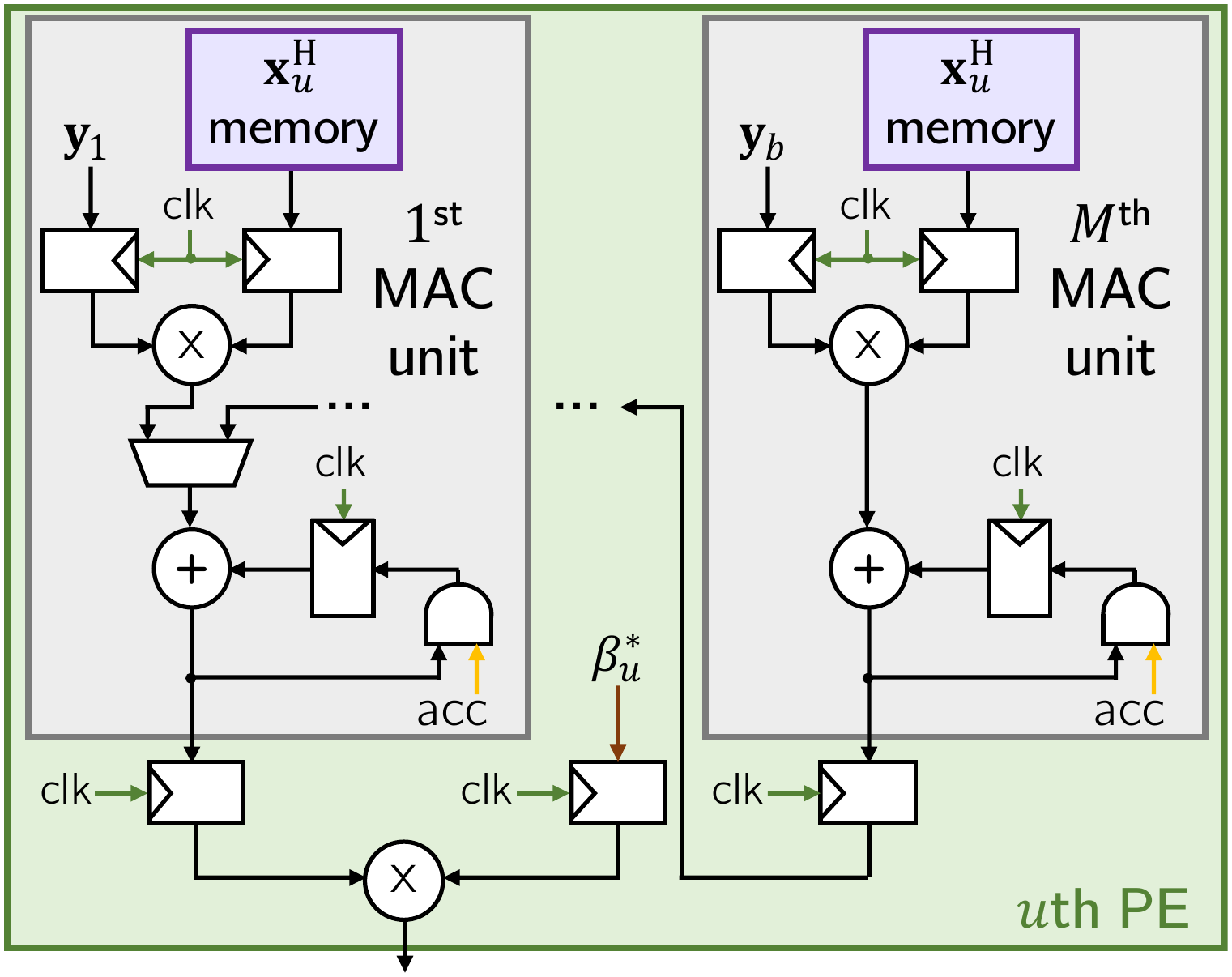}\\
\vspace{-0.2cm}
\caption{Optimized MAC array architecture. A processing element (PE) is used per UE and contains $M$ MAC units. Each of the $M$ MAC units operates on a different section of $\bmx_u^H$, which contains $B/M$ complex-valued numbers.}
\label{fig:mvp}
\end{figure}

The hardware efficiency of the baseline array of MAC units in~\cite{castaneda19fame} can be optimized by means of replication. 
As shown in \fref{fig:mvp}, we propose to use, for each UE, one processing element (PE) that consists of $M$ MAC units, each MAC unit having access to a $B/M$-dimensional partition of $\bmx_u^H$.
Then, each MAC unit within a PE receives different entries of $\bmy$ to compute the inner-product between its partitions of $\bmx_u^H$ and $\bmy$ in $B/M$ clock cycles.
The $M$ results are then merged together to complete the inner product $\bmx_u^H\bmy$.
This final reduction is achieved by reusing the adders in the MAC units following a binary-tree structure, which takes $\log_2 M$ clock cycles.
\subsection{PPAC: Parallel Processor in Associative CAM} 
\label{sec:ppac}
\begin{figure}[tp]
\centering
\subfigure[Multi-bit PPAC row]{\includegraphics[width=0.99\columnwidth]{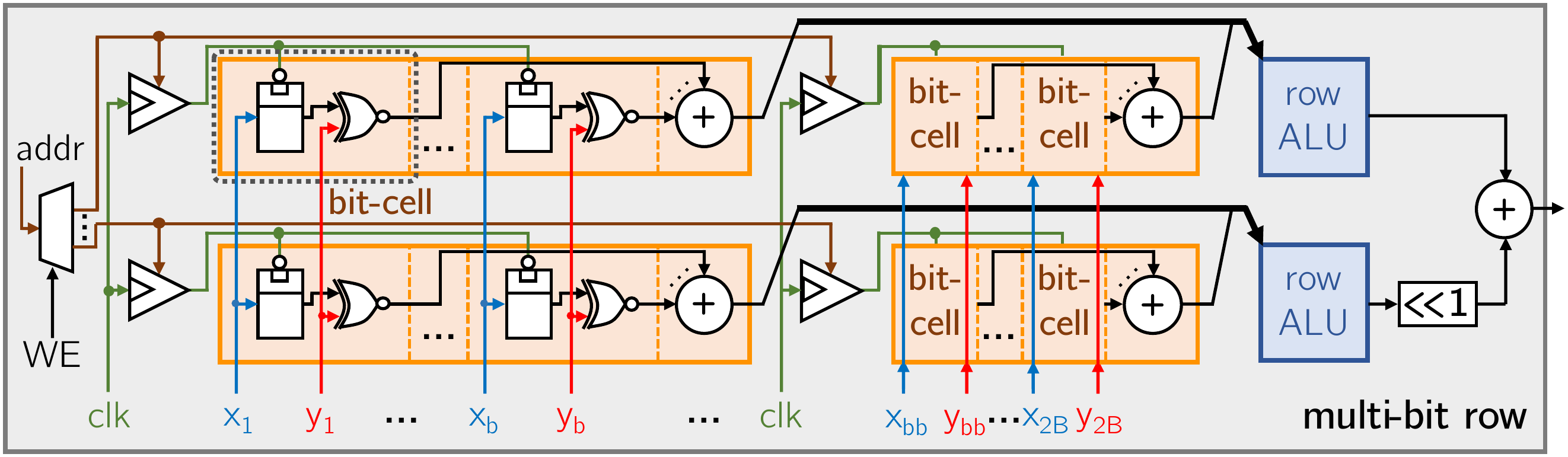}\label{fig:ppac_row}}
\subfigure[Row ALU details]{\includegraphics[width=0.33\columnwidth]{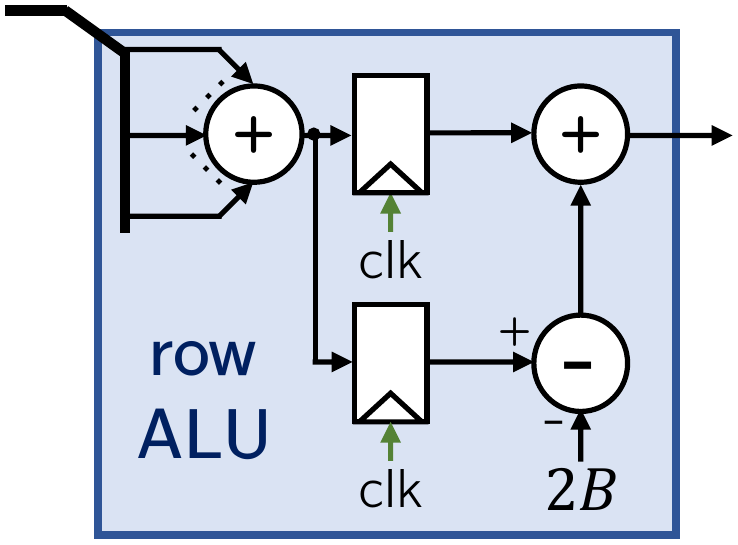}\label{fig:ppac_alu}}
\hspace{0.25cm}
\subfigure[PPAC PE]{\includegraphics[width=0.55\columnwidth]{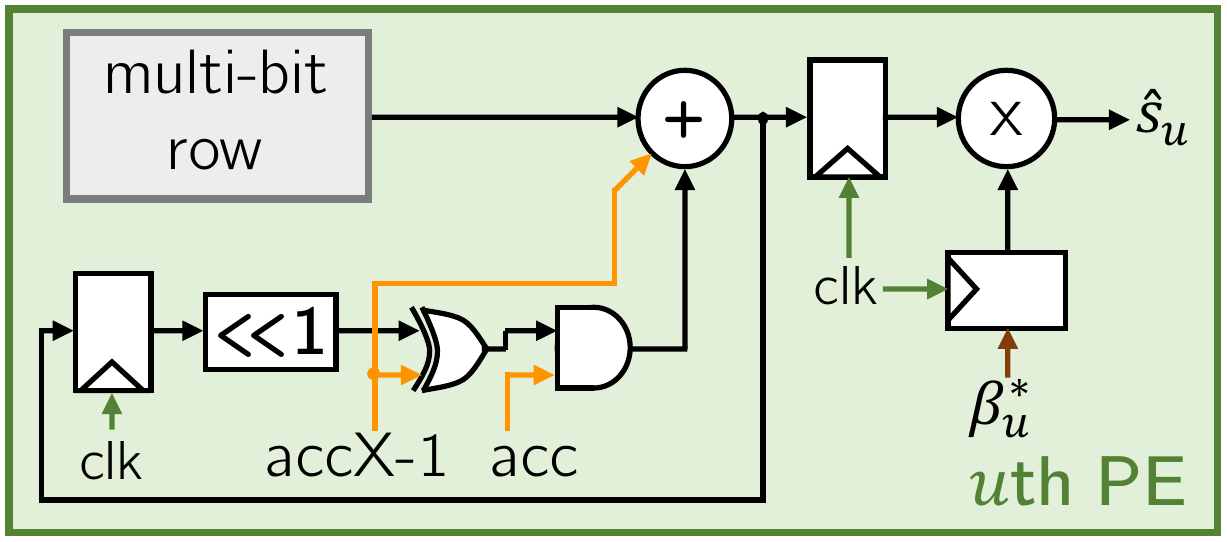}\label{fig:ppac_pe}}
\caption{Parallel Processor in Associative CAM (PPAC) specialized for finite-alphabet equalization. (a) Each bit-cell includes an XNOR that carries out a $1$-bit multiplication. For each matrix row, each bit-significance has its own PPAC row; the results are added with the proper scaling. (b) The row ALU sums the 1-bit products of a row and offsets them so that the result corresponds to an inner product. (c) Operation with an $L$-bit vector $\bmy$ is performed bit-serially across $L$ clock cycles. One processing element (PE) is used per UE.}
\label{fig:ppac}
\end{figure}
The PPAC architecture proposed in~\cite{castaneda2019ppac} is an all-digital CMOS PIM implementation in which every bit-cell of a memory is capable of multiplying its stored $1$-bit value with an external $1$-bit input using a bipolar (XNOR) or unipolar (AND) product. 
All products in a row are summed together by a population count in an arithmetic logic unit (ALU) associated with each row.
Each row ALU can further process the population count to accelerate a range of operations, including the execution of a $1$-bit matrix-vector product in a single clock cycle.
In this work, we simplify the PPAC architecture to the one illustrated in \fref{fig:ppac}, so that it only supports the matrix-vector products of interest for spatial equalization; see~\cite{castaneda2019ppac} for more details on PPAC.
\begin{table}[tp]
\renewcommand{\arraystretch}{1.1}
\begin{minipage}[c]{1\columnwidth}
    \centering
    \caption{Implementation results in 28\,nm CMOS for one PPAC equalizer instance operating in a system with  $B=256$, $U=16$, and $L=7$.}
    \label{tbl:implresultsoneppac}
  \begin{tabular}{@{}lccc@{}}
  \toprule
  $\bW^H$ resolution $K$ [bit] & $1$ & $2$ & $3$\\
  \midrule  
  {Silicon area [$\text{mm}^2$]} & 0.164 & 0.324 & 0.483 \\
  {Clock frequency [MHz]} & 796 & 785 & 784 \\
  {Throughput [M\,vectors/s]} & 113.7 & 112.1 & 112.0 \\
  {Power consumption\footnote{Extracted from stimuli-based post-layout simulations in the typical-typical process corner at $25^\circ\text{C}$ with a nominal power supply of 0.9V.} [mW]} & 112 & 246 & 383 \\
  \bottomrule
  \end{tabular}
  \end{minipage}
  \end{table}
\begin{table*}[tp]
\renewcommand{\arraystretch}{1.1}
\begin{minipage}[c]{2\columnwidth}
    \centering
    \caption{Implementation results in 28\,nm CMOS for various architectures operating at $2$~G\,vectors/s in a system with $B=256$ and $U=16$.}
    \label{tbl:implresults}
  \begin{tabular}{@{}l|ccc|ccc|ccc|ccc|ccc|ccc@{}}
  \toprule
  $\bmy$ resolution $L$ [bits] & \multicolumn{9}{c|}{4} & \multicolumn{9}{c}{7} \\
  Equalizer & \multicolumn{3}{c|}{Original MAC} & \multicolumn{3}{c|}{Optimized MAC} & \multicolumn{3}{c|}{PPAC} & \multicolumn{3}{c|}{Original MAC} & \multicolumn{3}{c|}{Optimized MAC} & \multicolumn{3}{c}{PPAC} \\
 $\bW^H$ resolution $K$ [bit] & $1$ & $2$ & $3$ & $1$ & $2$ & $3$ & $1$ & $2$ & $3$ & $1$ & $2$ & $3$ & $1$ & $2$ & $3$ & $1$ & $2$ & $3$\\
  \midrule  
  {Silicon area [$\text{mm}^2$]} & 21 & 32 & 42 & 4.1 & 7.0 & 8.6 & 1.8 & 3.6 & 5.3 & 22 & 33 & 43 & 5.6 & 9.8 & 11 & 3.0 & 5.8 & 8.7 \\
  {Power consumption\footnote{Extracted from stimuli-based post-layout simulations in the typical-typical process corner at $25^\circ\text{C}$ with a nominal power supply of 0.9V.} [W]} & 5.0 & 8.3 & 11.8 & 3.8 & 7.2 & 9.1 & 1.2 & 2.7 & 4.2 & 7.2 & 12.0 & 15.9 & 6.4 & 11.1 & 14.2 & 2.0 & 4.4 & 6.9\\
\bottomrule
\end{tabular}
\vspace{-0.2cm}
\end{minipage}
\end{table*}

As in~\cite{castaneda2019ppac}, our customized PPAC implementation supports multi-bit vectors in a bit-serial manner:
For an input vector $\bmy$ with $L$-bit entries, we first input a vector $\bmy_{[L]}$ which contains the most significant bit of all $\bmy$ entries, and we compute $\bX^H\bmy_{[L]}$, where $\bX^H$ is stored in the PPAC memory. This process is repeated $L$ times, while accumulating the new result to $2\times$ the previous result (cf.~\fref{fig:ppac_pe}) to compute $\bX^H\bmy$.
In contrast to~\cite{castaneda2019ppac}, we implement multi-bit matrix operations by having a different row for each bit-significance of a matrix entry:
If the matrix has $K$-bit entries, we use $K$ PPAC rows to represent one matrix row. The results coming from each row are combined using arithmetic shifts and additions (cf.~\fref{fig:ppac_row}).
Hence, each partial product $\bX^H\bmy_{[\ell]}$, $\ell=1,\ldots,L$, is computed in a single clock cycle, completing $\bX^H\bmy$ in $L$ clock cycles.

While PPAC naturally operates on real-valued numbers, spatial equalization requires complex-valued operations.
We use the real-valued decomposition of both $\bX^H$ and $\bmy$, i.e., we store $\bX^T_\reals$ in PPAC and apply $\bmy_\reals$ at the inputs, where
\begin{align*}
& \bX^T_\reals \!=\!  \!\left[\begin{array}{cc}
 \!\! \Re(\bX^H)  \!\!\!\! &~~-\Im(\bX^H)  \!\! \\
 \!\! \Im(\bX^H)  \!\!\!\! &~~\Re(\bX^H)  \!\!
\end{array}\right]\!,  \quad
& & \bmy_\reals \!=\! \!\left[\begin{array}{c}
 \!\! \Re(\bmy)  \!\! \\
 \!\! \Im(\bmy)   \!\! 
\end{array}\right]\!.
\end{align*}
As a result, the $U\times B$ complex-valued matrix-vector product between a matrix with $K$-bit entries and a vector with $L$-bit entries is computed in $L$ clock cycles using a PPAC array that has $2KU$ rows, each one with $2B$ bits.
The resulting $U$-dimensional vector $\bX^H\bmy$ is scaled with $\boldsymbol\beta^*$ using one complex-valued multiplier per UE~(cf.~\fref{fig:ppac_pe}), for a total of $U$ multipliers.
The markers in \fref{fig:ber} correspond to the fixed-point performance of our hardware design, which exhibits virtually no implementation loss when compared to double-precision floating-point arithmetic (represented by the curves).
\section{VLSI Design Comparison}
\subsection{Comparison Methodology}
We now compare VLSI implementation results for the different architectures described in \fref{sec:arch}.
We follow the procedure in~\cite{castaneda19fame}, where, for each equalizer resolution~$K$, a single instance of each VLSI architecture is placed-and-routed in $28$\,nm CMOS.
Power is measured only for matrix-vector products, not for initializing the memories.
\fref{tbl:implresultsoneppac} provides implementation results for PPAC. 
Since the throughput offered by a single PPAC instance is not sufficient to reach the throughputs targeted by future mmWave systems, we use several parallel instances in a time-interleaved manner to reach a target throughput of $2$\,G\,vectors/s. 
For example, for the case described in \fref{tbl:implresultsoneppac}, we need $18$ PPAC instances when operating with a 1-bit equalizer.
Then, we scale the area and power numbers proportionally to the number of instances.

The results for the linear array of MAC units are taken directly from~\cite{castaneda19fame}, while the ones for PPAC are obtained from our own VLSI designs.
The results for the optimized MAC array are estimated from the linear array of MAC units in the following way:
For a design with $K$-bit equalizer resolution, let $A_\text{MAC}$ be the fraction of silicon area occupied by the MAC units, excluding the $\bX^H$-memories, in the implementation results of the linear array of MAC units from~\cite{castaneda19fame}.
Then, $1-A_\text{MAC}$ corresponds to the fraction of the area occupied by the rest of the design, including the $\bX^H$-memories and $\boldsymbol\beta$-scaling multipliers.
As a result, the area occupied by the optimized MAC array with $M$ MAC units per PE can be modeled as $A=((1-A_\text{MAC})+M\times A_\text{MAC})$ times the area of the original array.
We follow the same procedure to estimate the power consumption of the optimized MAC array.
To compute the throughput of the optimized MAC array, we take the clock frequency of the original array from~\cite{castaneda19fame}, and model the latency of the optimized array with $T=B/M+\log_2(M)$ clock cycles, as described in \fref{sec:repmac}.
We choose the parameter $M$ that minimizes the $AT$-product while being a power of two to perform reduction as described for the optimized MAC array.

\subsection{Comparison of Finite-Alphabet Equalizers}

\fref{tbl:implresults} shows implementation results for a system with $B=256$ antennas, $U=16$ UEs, and for a target throughput of $2$\,G\,vectors/s for the cases where the equalizer $\bW^H$ resolution is $K \in \{1,2,3\}$ bits and the received vector $\bmy$ resolution is $L \in \{4,7\}$ bits.
For the optimized MAC array, we use $M=16$ MAC units per UE.
We observe that the optimized MAC array improves the circuit area of the original array by more than $3\times$; this is because the MAC units occupy no more than $20\%$ of the original design's area, so that replication does not increase the per-instance area significantly, but has a significant impact on the per-instance throughput, requiring fewer instances to meet the target throughput. 
While there are fewer instances, each instance contains more MAC units, which draw up to $75\%$ of the initial design's power.
As a result, both MAC arrays consume roughly the same amount of power. 

By comparing the optimized MAC array results to those of PPAC, we observe that PPAC achieves a significant reduction in silicon area and power consumption for all the considered cases.
For example, when considering $1$-bit finite-alphabet equalizers with $4$-bit inputs, we see that PPAC reduces the area and power consumption by a factor of $2.3\times$ and $3.1\times$, respectively.
However, these area and power consumption savings decrease as the equalizer resolution increases: When using a $3$-bit equalizer with $4$-bit inputs, PPAC offers $1.6\times$ and $2.2\times$ lower area and power consumption, respectively.
These results illustrate that bit-serial PIM architectures provide benefits not only in scenarios that need a massive amount of memory, but also in applications that require hardware implementations with low area and power, where ASICs have been used traditionally.
We finally emphasize  that the theoretical advantages of PIM can already be realized with an all-digital architecture implemented in standard CMOS technology, as demonstrated by our PPAC results.
\section{Conclusions}
We have implemented finite-alphabet equalization for mmWave massive MU-MIMO systems using traditional ASIC architectures and a PIM architecture. 
Our implementation results have shown that an all-digital PIM solution reduces the silicon area and power consumption by at least a factor of $1.2\times$ and $2.1\times$, respectively, compared to a conventional ASIC.
Combined with the savings provided by finite-alphabet equalization (up to $5.8\times$ and $3.9\times$ reduction in area and power consumption, respectively), PIM architectures pave the way for high-throughput and low-power all-digital BS architectures.

There are many avenues for future work. A system-level analysis that studies the effects of the different input interfaces used by the considered architectures is in order. Furthermore, other PIM designs that use, e.g., mixed-signal techniques or that rely on emerging devices, should be considered, as they could achieve superior throughput, area, and power consumption. 

\balance


\begin{thebibliography}{10}
\providecommand{\url}[1]{#1}
\csname url@samestyle\endcsname
\providecommand{\newblock}{\relax}
\providecommand{\bibinfo}[2]{#2}
\providecommand{\BIBentrySTDinterwordspacing}{\spaceskip=0pt\relax}
\providecommand{\BIBentryALTinterwordstretchfactor}{4}
\providecommand{\BIBentryALTinterwordspacing}{\spaceskip=\fontdimen2\font plus
\BIBentryALTinterwordstretchfactor\fontdimen3\font minus
  \fontdimen4\font\relax}
\providecommand{\BIBforeignlanguage}[2]{{%
\expandafter\ifx\csname l@#1\endcsname\relax
\typeout{** WARNING: IEEEtran.bst: No hyphenation pattern has been}%
\typeout{** loaded for the language `#1'. Using the pattern for}%
\typeout{** the default language instead.}%
\else
\language=\csname l@#1\endcsname
\fi
#2}}
\providecommand{\BIBdecl}{\relax}
\BIBdecl

\bibitem{swindlehurst14a}
A.~L. Swindlehurst, E.~Ayanoglu, P.~Heydari, and F.~Capolino, ``Millimeter-wave
  massive {MIMO}: The next wireless revolution?'' \emph{{IEEE} Commun. Mag.},
  vol.~52, no.~9, pp. 56--62, Sep. 2014.

\bibitem{larsson14a}
E.~G. Larsson, F.~Tufvesson, O.~Edfors, and T.~L. Marzetta, ``Massive {MIMO}
  for next generation wireless systems,'' \emph{{IEEE} Commun. Mag.}, vol.~52,
  no.~2, pp. 186--195, Feb. 2014.

\bibitem{roh2014millimeter}
W.~Roh, J.-Y. Seol, J.~Park, B.~Lee, J.~Lee, Y.~Kim, J.~Cho, K.~Cheun, and
  F.~Aryanfar, ``Millimeter-wave beamforming as an enabling technology for {5G}
  cellular communications: Theoretical feasibility and prototype results,''
  \emph{{IEEE} Commun. Mag.}, vol.~52, no.~2, pp. 106--113, Feb. 2014.

\bibitem{sadhu20177}
B.~Sadhu, Y.~Tousi, J.~Hallin, S.~Sahl, S.~Reynolds, {\"O}.~Renstr{\"o}m,
  K.~Sj{\"o}gren, O.~Haapalahti, N.~Mazor, B.~Bokinge, G.~{Weibull},
  H.~{Bengtsson}, A.~{Carlinger}, E.~{Westesson}, J.~{Thillberg}, L.~{Rexberg},
  M.~{Yeck}, X.~{Gu}, D.~{Friedman}, and A.~{Valdes-Garcia}, ``A 28{GHz}
  32-element phased-array transceiver {IC} with concurrent dual polarized beams
  and 1.4 degree beam-steering resolution for {5G} communication,'' in
  \emph{IEEE Int. Solid-State Circuits Conf. (ISSCC)}, Feb. 2017, pp. 128--129.

\bibitem{du2018hybrid}
J.~Du, W.~Xu, H.~Shen, X.~Dong, and C.~Zhao, ``Hybrid precoding architecture
  for massive multiuser {MIMO} with dissipation: Sub-connected or fully
  connected structures?'' \emph{{IEEE} Trans. Wireless Commun.}, vol.~17,
  no.~8, pp. 5465--5479, Aug. 2018.

\bibitem{magueta2019hybrid}
R.~Magueta, D.~Castanheira, A.~Silva, R.~Dinis, and A.~Gameiro, ``Hybrid
  multi-user equalizer for massive {MIMO} millimeter-wave dynamic subconnected
  architecture,'' \emph{{IEEE} Access}, vol.~7, pp. 79\,017--79\,029, 2019.

\bibitem{alkhateeb14b}
A.~Alkhateeb, J.~Mo, N.~Gonz{\'a}lez-Prelcic, and R.~W. {Heath Jr.}, ``{MIMO}
  precoding and combining solutions for millimeter-wave systems,'' \emph{{IEEE}
  Commun. Mag.}, vol.~52, no.~12, pp. 122--131, Dec. 2014.

\bibitem{bjornson2019massive}
E.~Bj{\"o}rnson, L.~Van~der Perre, S.~Buzzi, and E.~G. Larsson, ``Massive
  {MIMO} in sub-6 {GHz} and {mmWave}: Physical, practical, and use-case
  differences,'' \emph{{IEEE} Wireless Commun.}, vol.~26, no.~2, pp. 100--108,
  Apr. 2019.

\bibitem{dutta2019case}
S.~Dutta, C.~N. Barati, A.~Dhananjay, D.~A. Ramirez, J.~F. Buckwalter, and
  S.~Rangan, ``A case for digital beamforming at {mmWave},'' \emph{{IEEE}
  Trans. Wireless Commun.}, vol.~19, no.~2, pp. 756--770, Feb. 2020.

\bibitem{mo15b}
J.~Mo and R.~W. {Heath Jr.}, ``Capacity analysis of one-bit quantized {MIMO}
  systems with transmitter channel state information,'' \emph{{IEEE} Trans.
  Signal Process.}, vol.~63, no.~20, pp. 5498--5512, Oct. 2015.

\bibitem{roth2017achievable}
K.~Roth and J.~A. Nossek, ``Achievable rate and energy efficiency of hybrid and
  digital beamforming receivers with low resolution {ADC},'' \emph{{IEEE} J.
  Sel. Areas Commun.}, vol.~35, no.~9, pp. 2056--2068, Sep. 2017.

\bibitem{jacobsson17b}
S.~Jacobsson, G.~Durisi, M.~Coldrey, U.~Gustavsson, and C.~Studer, ``Throughput
  analysis of massive {MIMO} uplink with low-resolution {ADCs},'' \emph{{IEEE}
  Trans. Wireless Commun.}, vol.~16, no.~6, pp. 4038--4051, Jun. 2017.

\bibitem{castaneda19fame}
O.~Casta{\~n}eda, S.~Jacobsson, G.~Durisi, T.~Goldstein, and C.~Studer,
  ``Finite-alphabet {MMSE} equalization for all-digital massive {MU-MIMO
  mmWave} communication,'' \emph{{IEEE} J. Sel. Areas Commun.}, vol. 38, no.9, pp. 2128--2141, Sep. 2020.

\bibitem{studer16a}
C.~Studer and G.~Durisi, ``Quantized massive {MU-MIMO-OFDM} uplink,''
  \emph{{IEEE} Trans. Commun.}, vol.~64, no.~6, pp. 2387--2399, Jun. 2016.

\bibitem{yan2019performance}
H.~Yan, S.~Ramesh, T.~Gallagher, C.~Ling, and D.~Cabric, ``Performance, power,
  and area design trade-offs in millimeter-wave transmitter beamforming
  architectures,'' \emph{{IEEE} Circuits Syst. Mag.}, vol.~19, no.~2, pp.
  33--58, May 2019.

\bibitem{studer2011asic}
C.~Studer, S.~Fateh, and D.~Seethaler, ``{ASIC} implementation of soft-input
  soft-output {MIMO} detection using {MMSE} parallel interference
  cancellation,'' \emph{{IEEE} J. Solid-State Circuits}, vol.~46, no.~7, pp.
  1754--1765, Jul. 2011.

\bibitem{wu2014large}
M.~Wu, B.~Yin, G.~Wang, C.~Dick, J.~R. Cavallaro, and C.~Studer, ``Large-scale
  {MIMO} detection for {3GPP} {LTE}: Algorithms and {FPGA} implementations,''
  \emph{{IEEE} J. Sel. Topics Signal Process.}, vol.~8, no.~5, pp. 916--929,
  Oct. 2014.

\bibitem{wulf1995hitting}
W.~A. Wulf and S.~A. McKee, ``Hitting the memory wall: Implications of the
  obvious,'' \emph{ACM SIGARCH Comput. Archit. News}, vol.~23, no.~1, pp.
  20--24, Mar. 1995.

\bibitem{nair2015evolution}
R.~Nair, ``Evolution of memory architecture,'' \emph{Proc. {IEEE}}, vol. 103,
  no.~8, pp. 1331--1345, Aug. 2015.

\bibitem{li2017drisa}
S.~Li, D.~Niu, K.~T. Malladi, H.~Zheng, B.~Brennan, and Y.~Xie, ``{DRISA}: A
  {DRAM}-based reconfigurable in-situ accelerator,'' in \emph{IEEE/ACM Int.
  Symp. Microarchit. (MICRO)}, Oct. 2017, pp. 288--301.

\bibitem{jia2018microprocessor}
H.~Jia, Y.~Tang, H.~Valavi, J.~Zhang, and N.~Verma, ``A microprocessor
  implemented in 65nm {CMOS} with configurable and bit-scalable accelerator for
  programmable in-memory computing,'' \emph{arXiv preprint arXiv:1811.04047},
  Nov. 2018.

\bibitem{guo2013ac}
Q.~Guo, X.~Guo, R.~Patel, E.~Ipek, and E.~G. Friedman, ``{AC-DIMM}: Associative
  computing with {STT-MRAM},'' \emph{ACM SIGARCH Comput. Archit. News},
  vol.~41, no.~3, pp. 189--200, Jun. 2013.

\bibitem{castaneda2019ppac}
O.~Casta{\~n}eda, M.~Bobbett, A.~Gallyas-Sanhueza, and C.~Studer, ``{PPAC}: A
  versatile in-memory accelerator for matrix-vector-product-like operations,''
  in \emph{IEEE Int. Conf. Appl.-specific Syst., Archit. and Processors
  (ASAP)}, Jul. 2019, pp. 149--156.

\bibitem{paulraj03}
A.~Paulraj, R.~Nabar, and D.~Gore, \emph{Introduction to space-time wireless
  communications}.\hskip 1em plus 0.5em minus 0.4em\relax Cambridge Univ.
  Press, 2003.

\bibitem{castaneda20a}
O.~Casta\~{n}eda, S.~Jacobsson, G.~Durisi, T.~Goldstein, and C.~Studer,
  ``Soft-output finite alphabet equalization for {mmWave} massive {MIMO},'' in
  \emph{Proc. IEEE Int. Conf. Acoust., Speech, Signal Process. (ICASSP)}, May
  2020, pp. 1763--1767.

\bibitem{jaeckel2014quadriga}
S.~Jaeckel, L.~Raschkowski, K.~B{\"o}rner, and L.~Thiele, ``{QuaDRiGa}: A {3-D}
  multi-cell channel model with time evolution for enabling virtual field
  trials,'' \emph{{IEEE} Trans. Antennas Propag.}, vol.~62, no.~6, pp.
  3242--3256, Jun. 2014.

\end{thebibliography}
\end{document}